\documentclass[12pt,letterpaper]{article}
\usepackage[sort&compress, numbers]{natbib}
\usepackage{graphicx}
\usepackage{amssymb}
 \usepackage[margin=1in,top=0.5in,bottom=0.6in]{geometry}
\usepackage{lineno}
\usepackage{enumitem}
\usepackage{graphics}
\usepackage{rotating}
\usepackage{hyperref}
\usepackage{aas_macros}
\newcommand{\msun}{M_{\odot}}

\usepackage{natbib}
\usepackage{verbatim}
\usepackage{titlesec}

\begin{document}

\titlespacing{\section}{0pt}{\parskip}{0pt}
\titlespacing{\subsection}{0pt}{\parskip}{-\parskip}

\begin{titlepage}
	\begin{center}
	{\large\bfseries Astro2020 Science White Paper\par}
	\vspace{0.6cm}
	{\scshape\LARGE 
 Gravitational Waves, Extreme Astrophysics, and Fundamental Physics with Precision  Pulsar Timing  }
	\vspace{0.5cm}\\
	{\large \textbf{Principal authors}} \linebreak
    {\large {\scshape James Cordes}} {\normalsize\textit{(Cornell University)}}, {\normalsize\href{mailto:jcordes@astro.cornell.edu}{cordes@astro.cornell.edu}}
    	\linebreak
	{\large {\scshape\large Maura McLaughlin}} {\normalsize\textit{(West Virginia University)}}, {\normalsize\href{mailto:maura.mclaughlin@mail.wvu.edu}{maura.mclaughlin@mail.wvu.edu}} \par

	\vspace{0.2cm}
	{\scshape\Large for the NANOGrav Collaboration\par}
	{\it \large (The North American Nanohertz Observatory for Gravitational Waves)}

\begin{figure}[h]
    \centering
    \includegraphics[width=\linewidth]{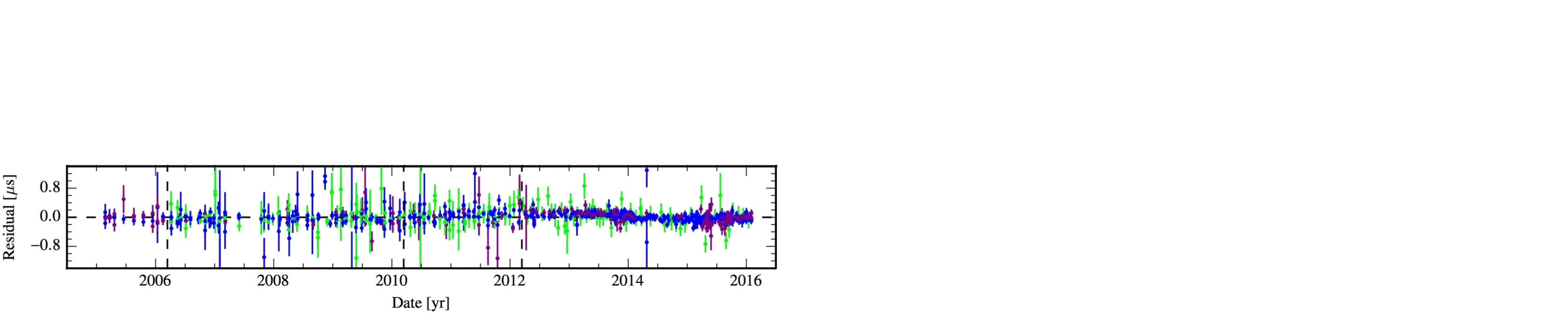}
    \caption{\footnotesize Over a decade of timing residuals from millisecond pulsar J1713+0747 with the Arecibo Observtory and Green Bank Telescope at 820 MHz (green), 1.4 GHz (blue), and 2.1 GHz (magenta) \cite{2018ApJS..235...37A}. The RMS residual  is 116~ns, demonstrating the remarkable stability of millisecond pulsars over long timespans. }
    \label{fig:1713resid}
\end{figure}

	\vspace{0.5cm}
		{\normalsize This is one of five core white papers written by members of the NANOGrav Collaboration.} 
		
\end{center}
        \noindent
{\it  Supermassive Black-hole Demographics \& Environments with Pulsar Timing Arrays},  \\ S.  Taylor et al.\\
\smallskip
        \noindent
    {\it Fundamental Physics with Radio Millisecond  Pulsars}, E. Fonseca et al.\\
    \smallskip
        \noindent
    {\it Physics Beyond the Standard Model with Pulsar Timing Arrays}, X. Siemens et al. \\
    \smallskip
        \noindent {\it Multi-messenger Astrophysics with Pulsar Timing Arrays}, L.~Kelley et al.
        \noindent
        \center{\scshape\large Other Related Science Whitepapers}

{\it \noindent Gravitational-Wave Astronomy in the 2020s and Beyond: A view across the gravitational wave spectrum}, the Gravitational Waves International Committee\\
{\it \noindent The Virtues of Time and Cadence for Pulsars and Fast Transients}, R. Lynch et al.\\
    {\it  \noindent Twelve Decades: Probing the ISM from kiloparsec to sub-AU scales}, D. Stinebring et al.
    \vspace{0.5in}

	\noindent \textbf{Thematic Areas:} \hspace*{60pt} 
	$\square$ Planetary Systems \hspace*{10pt} 
	$\square$ Star and Planet Formation \hspace*{10pt}\linebreak
    $\square$ Formation and Evolution of Compact Objects \hspace*{10pt} 
    ${\rlap{$\checkmark$}}\square$ Cosmology and Fundamental Physics \linebreak
    $\square$  Stars and Stellar Evolution \hspace*{1pt} $\square$ Resolved Stellar Populations and their Environments \hspace*{40pt} 
     $\square$    Galaxy Evolution   \hspace*{45pt} 
     ${\rlap{$\checkmark$}}\square$             Multi-Messenger Astronomy and Astrophysics \hspace*{65pt} \linebreak
		\vfill
    
		\vfill
\end{titlepage}

\section{\large Probing the Universe with Precision Timing }
Precision pulsar timing at the level of tens to hundreds of nanoseconds allows detection of nanohertz gravitational waves (GWs) from  supermassive binary black holes (SMBBHs) at the cores of merging galaxies and, potentially, from exotic sources such as cosmic strings.  The same timing  data used for GW astronomy also yield precision masses of neutron stars orbiting other  compact objects, constraints on the equation of state of nuclear matter,  and precision tests of General Relativity, the Strong Equivalence Principle, and alternative theories of gravity. Timing can also lead to stringent constraints on the photon mass and on changes in fundamental constants and could reveal low mass objects (rogue planets,  dark matter clumps) that traverse  pulsar lines of sight. Data sets also allow modeling of the density, magnetic field, and  turbulence  in the interstellar plasma.
Roughly 100 millisecond pulsars (MSPs) are currently being timed at $\sim$GHz frequencies using the largest radio telescopes in the world. These
 data sets currently represent $\sim$1000 pulsar-years and will increase dramatically over the next decade.  These topics are presented  as a program of key science with flowdown technical requirements for achieving the science.

\section{\large Pulsar Timing}
 The time of arrival  of astrophysical pulses is determined by the path of the pulse, and the intervening magnetized plasma that invariably exists between source and observer.

The precision of a  time of  arrival (TOA)  depends  on the pulse width  and 
on the signal-to-noise ratio, thus favoring objects that emit bright, narrow pulses and observations using large telescopes with sensitive, wide-bandwidth receivers.  At present the best  TOA precisions of tens of ns are achieved with 100-m class or larger telescopes at $\sim$GHz frequencies and 
 GHz bandwidths (see Fig.~\ref{fig:1713resid} and \cite{2018ApJS..235...37A}). These precisions can be improved for a large number of MSPs using a variety of methods we discuss below.

GWs and other phenomena produce TOA departures from a comprehensive model that is constructed  for each pulsar. These models  account for all known effects, including the spin, spindown, position, proper motion, and parallax of the pulsar,
along with those extrinsic to the pulsar, such as any binary motion, the movement of our local reference frame, and  frequency-dependent delays from  the interstellar plasma.  GW detection requires sophisticated noise models that include astrophysical effects and measurement errors.
\citep[][]{2013CQGra..30v4002C}.  The primary astrophysical effects are pulse jitter intrinsic to the pulsar \cite{cd85,FDJitter}, low-level noise in the pulsar spin \cite{sc10,NG9EN}, and imperfect removal of chromatic interstellar propagation effects \cite{cs10,LamDMt,optimalfreq,sc17}.   
\newcommand{\sigR}{\sigma_{\cal R}}
The accuracy of timing models is 
 quantified by the RMS TOA residuals (measured minus predicted arrival times), which include TOA errors and other contributions. 
RMS values  $\sigR < 100$~ns over timespans of decades have been achieved   \cite{2015ApJ...809...41Z,2018ApJS..235...37A}.

Since the first MSP discovery in 1982, radio surveys  have revealed about 300  in our Galaxy and another 150 in globular clusters \cite{2005AJ....129.1993M}.
Ensembles of Galactic MSPs, or pulsar timing arrays (PTAs), are being observed with telescopes around the world to search for the TOA perturbations  expected from nanohertz GWs. Roughly 100 MSPs are currently included in this effort and this number continues to grow as more bright MSPs are found. These datasets also allow a wide variety of other fundamental measurements, including the masses of compact objects (neutron stars, white dwarfs) in binary systems, precision tests of General Relativity,  photon mass limits, and constraints on dark matter clumps and cosmic strings.

In the next decade, increased access to large telescopes along with  algorithmic advances will lead to discoveries in all of these science areas.

\section{\large Detection and Characterization of Nanohertz Gravitational Waves}

\newcommand{\fgw}{f_{\rm GW}}

To exploit MSPs as astrophysical clocks, they have been assembled into  Galactic-scale detectors (PTAs) of long-wavelength gravitational waves. 
PTAs can detect strains  $h \sim \sigR/T$ at frequencies where they are most sensitive, $\fgw \sim 1/T$, where $T$ is the total timespan of observations.
Decade-long timing data sets with  $\sigR \sim 100$~ns therefore probe strains of $10^{-15}$ at frequencies $\sim 10^{-8}$ to $10^{-9}$~Hz.  Detection and identification of GWs is
 aided by the angular correlation  of timing residuals between pairs of pulsars. 
 The angular correlation function is known for General Relativity \cite{1983ApJ...265L..39H, 2013PhRvD..88f2005M,2014PhRvD..90f2011M, 2018JPhCo...2j5002M,2014PhRvD..90h2001G} and for alternate theories of gravity \cite{1983ApJ...265L..39H,2008ApJ...685.1304L,2012PhRvD..85h2001C,2015PhRvD..92j2003G}.

\begin{figure}
    \centering
    \includegraphics[width=0.6\linewidth]{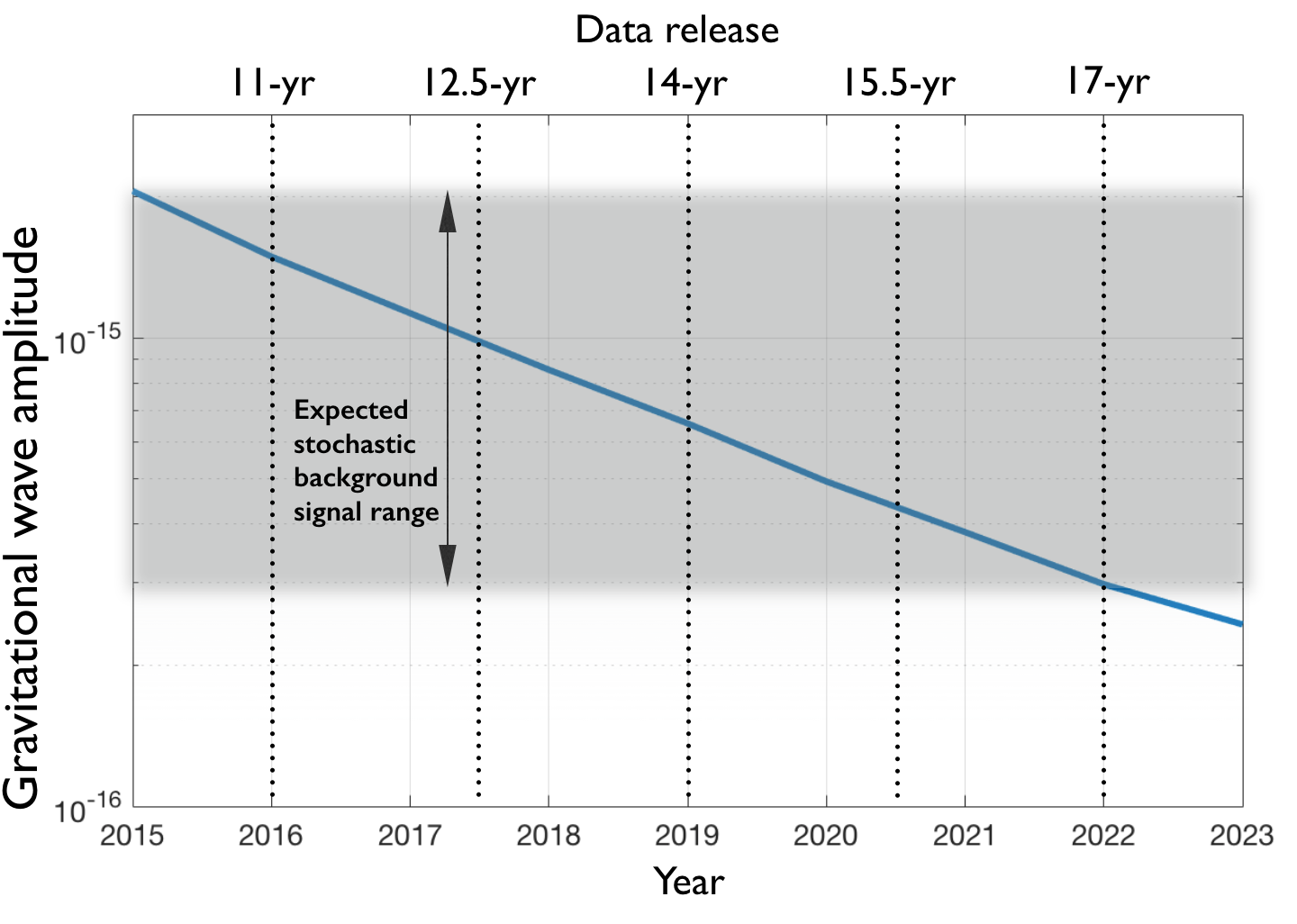}
    \vspace{-0.15in}
    \caption{\footnotesize Detectable amplitudes of a stochastic background of GWs produced by SMBBH mergers as a function of time  
    \cite{2018ApJS..235...37A,2018ApJ...859...47A}. The blue curve corresponds to  the 95\% upper limit expected to be set by NANOGrav in the absence of a GW signal. The shaded region corresponds to the 1-$\sigma$ range of recent background estimates produced using a range of galaxy merger rates and empirical black hole-host relations \cite{2014ApJ...789..156M,2016ApJ...826...11S,2016MNRAS.463L...6S}. The dotted  lines show the cut-off times of  previous and planned NANOGrav data releases. }
    \label{fig:sbdetection}
\end{figure}

Galaxy mergers form the backbone of current structure-formation models \cite{2015ARA&A..53...51S}. As a byproduct,  their central supermassive black holes form SMBBHs that eventually merge due to GW emission in the nanohertz regime.  
These comprise the dominant expected sources of 
low-frequency GWs.   They may be detected individually  as continuous wave sources or as burst sources from the memory effect at the time of merger of the SMBBH.   However, the collective {\it  stochastic background} from the incoherent superposition of GWs from the cosmic history of black hole mergers   is likely to be detected first with dimensionless strain $h \sim10^{-15}$.   
The stochastic background is expected to have a steep spectrum that increases at lower frequencies.  TOA perturbations therefore grow  
$\propto T^{5/3}$ as longer data spans sample increasingly lower frequencies \cite{2013CQGra..30v4015S}.
Long, continuous data sets obtained with stable instrumentation are  a {\it fundamental} requirement for nHz GW science.

\begin{figure}
    \centering
    \includegraphics[width=0.7\linewidth]{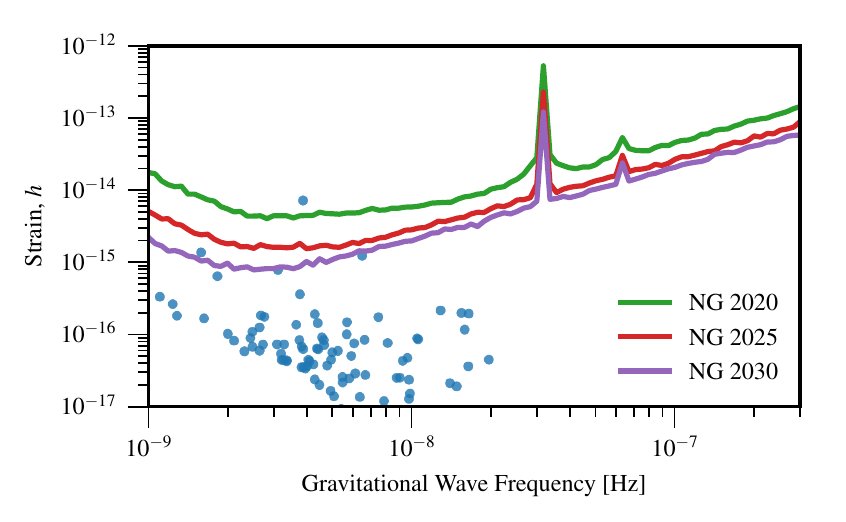}
    \vspace {-0.15in}
    \caption{\footnotesize  Example realization of the local GW sky, with  nanohertz sources as blue dots. The curves are $95\%$ detection probability  for NANOGrav in 2020, 2025, and 2030, with a $3\times10^{-3}$ false alarm rate \cite{2017NatAs...1..886M}. Averaging over  realizations,  one to several individual SMBBHs should be detectable within the next decade.}
    \label{fig:cwdetection}
\end{figure}

With modest projected increases in current sensitivity in the near term as  additional  MSPs are included,  
the stochastic background is expected to be detected
within the next several years \citep{2016ApJ...819L...6T} (Fig.~\ref{fig:sbdetection}) and GWs from individual SMBH binaries before the end of the next decade \cite{2017NatAs...1..886M} (Fig.~\ref{fig:cwdetection}).

 Other possible GW sources include cosmic strings and primordial GWs (see white paper {\it Physics Beyond the Standard Model with Pulsar Timing Arrays}, Siemens et al.).
The most sensitive limits on the stochastic GW background already inform models for galaxy evolution, disfavoring those with large black hole to galactic bulge mass ratios  and placing constraints on the astrophysical processes involved in galaxy mergers, demonstrating that they are not solely driven by gravitational wave emission but also through  processes such as stellar hardening, eccentricity, and gas dynamics \cite{2018ApJ...859...47A,2017PhRvL.118r1102T,2018arXiv181108826B,2019NatAs...3....8M}
(see white paper {\it Supermassive Black-hole Demographics \& Environments with Pulsar Timing Arrays}, Taylor et al.).
They also set the most stringent limits on the properties of cosmic strings 
\citep[][]{2018ApJ...859...47A} and can be used to place sensitive limits on individual SMBBH sources in local group galaxies \cite{2018arXiv181211585A}.

PTAs are also sensitive to Galactic sources of GWs and to small masses that can perturb the location of the pulsar or the solar system or that perturb arrival times directly through gravitational lensing or the Shapiro delay \cite{2007ApJ...659L..33S,2011PhRvD..84d3511B,2012MNRAS.426.1369K,2019arXiv190104490D}.  Bursts with memory  are potentially detectable along lines of sight to pulsars close to any merging binaries, such as in the Galactic center or in globular clusters
\citep[][]{2005ApJ...627L.125J,2012ApJ...752...67K,2017PhRvD..96l3016M}.

\section{\large Understanding the Physics of Pulsar Spins}

\newcommand{\Msun}{M_{\odot}}

The rapid spins of MSPs comprise the astrophysical clocks that underlie their
use for gravitational wave detection.  Most pulsars, including some MSPs, show 
 spin fluctuations that contribute red noise 
that limits GW sensitivity at very low ($\lesssim 0.1$~cy~yr$^{-1}$) frequencies.    Red noise is  much weaker in MSPs compared to other pulsars, but its  physics is not well understood.   Longer data sets on more MSPs will allow better empirical
characterization  that will further optimize GW detection.  Development of a better scaling law of the RMS red noise vs. spin rate, magnetic field, and other parameters will allow better forecasting of timing quality  for  selection of  MSPs for inclusion in PTAs.   Masses of MSPs, which  vary  from $\sim 1.3$ to $2~\Msun$, may be one of the `hidden' variables in the scaling law 
\citep[][]{sc10}.
 Increasing the MSP mass measurement sample to 100 or more   is key to exploring this further.

\section{\large Fundamental Physics}

Timing of pulsars in double neutron star (DNS) systems has famously provided tests of General Relativity to better than 0.01\% precision \cite{2008ARA&A..46..541K,2016ApJ...829...55W}.  More extreme binary systems remain to be discovered (e.g. DNS systems with sub-hour orbits,  stellar mass black hole-pulsar systems, pulsars orbiting Sgr~A*); they will provide fundamentally stronger constraints on alternative theories of gravity and on the environment of the Galactic center. 
Post-Newtonian terms in relativistic binaries also yield precision masses of neutron stars and their companions that have ruled out many equations of state for supranuclear matter \cite{2010Natur.467.1081D,2013Sci...340..448A}. Timing observations of a triple system (a pulsar with two white dwarf companions) have placed the best constraints yet on the Strong Equivalence Principle \cite{2018Natur.559...73A}.
Alternative theories of gravity also impact the properties of gravitational waves and their timing perturbations.  The angular correlation, for example, is altered if there are more than two GW polarizations \cite{2008ApJ...685.1304L}
(see whitepaper {\it Fundamental Physics with Radio Millisecond  Pulsars}, Fonseca et al.).

\section{\large Exotic Physics}

Numerous low-mass objects in the Galaxy can produce detectable gravitational perturbations on an individual pulsar or on the Solar System and thus a collective effect on all lines of sight. 
  Gravitational lensing from passage close to pulsar line of sight is also possible, with the small optical depth  requiring many pulsar-years of timing measurements.  Dark matter clumps on mass scales of $10^{-11}$ to $10^{-8}~\msun$ \citep[][]{2014JPhG...41f3101R} are potentially detectable after a few thousand pulsar-years of high-precision ($\sigR\sim 10s$ of ns) timing \cite{2018arXiv180107847K}, achievable in a long-term project. 

Warm or ``fuzzy'' dark matter \cite{2000PhRvL..85.1158H,2017PhRvD..95d3541H} may also cause oscillations of the Galactic gravitational potential, which would manifest as low-frequency gravitational waves; PTA data currently sets the tightest constraints on the presence of ultralight bosons
\cite{2018PhRvD..98j2002P}.
 Timing of binary MSPs can also be used to place constraints on a variable gravitational constant and MSP spin-down measurements may also be used to constrain the mass of the photon
 (see white paper {\it Physics Beyond the Standard Model with Pulsar Timing Arrays},
 Siemens et al.).

\section{\large Science Requirements}

The full science program requires suitable pulsars, including those currently known and the anticipated larger number of new MSPs and relativistic binaries that will be discovered next decade.    The program also requires sustained observations on stable instrumentation.  In particular, 
PTA science requires observations of 100 to 200  MSPs at regular cadence over time spans of decades. Uniform sky coverage is 
required to probe the detailed spectrum and isotropy of the stochastic background and to detect individual sources.  Uniform GW sensitivity over the sky necessarily requires multiple telescopes for the timing program and it is essential to have overlapping coverage on  a few objects for calibration purposes.     {\it This will require increased access to high-sensitivity telescopes beginning as soon as possible.  }

{\bf Improving  Sensitivity to GWs and Post-Newtonian Effects:}

{\bf   Longer data sets.}
Precision MSP timing data sets using modern instrumentation currently span one to two decades, but increasingly longer data sets will have transformative effects on PTA science. The signals from the stochastic GW background and from bursts with memory grow as $\sim T^{5/3}$ and $T$, respectively. A crucial point is that errors in the solar system ephemeris decouple from GW signals for data sets longer than the orbital periods of Jupiter and Saturn (which are comparable to current data set lengths), providing greater sensitivity to GWs. Further benefits of long datasets include increased sensitivity to small bodies in the Solar System 
\citep[][]{2018MNRAS.481.5501C}
or Milky Way, and to dark matter substructure \citep[][]{2007MNRAS.382..879S}.

{\bf Greater collecting area.}  Timing precision of the majority of MSPs in  PTAs  is limited by radiometer noise 
\citep[][]{2018ApJS..235...37A}. Noise modeling indicates
that  telescopes with sensitivities up to twice that of Arecibo would  significantly improve TOAs for most currently timed pulsars and would allow many more objects to be used.  

{\bf Higher observational cadences and wider bandwidths.}  Higher cadence (i.e. daily--weekly) and wide-bandwidth observations will increase TOA precision via better 
mitigation of  chromatic interstellar delays
\cite{2018ApJS..235...37A}. Faster sampling also increases the probability of identifying rapid gravitational effects from dark matter clumps. 

{\bf Larger numbers of MSPs.}  A PTA's sensitivity to GWs increases linearly with the number of MSPs. For this reason, and to achieve  uniform GW sensitivity across the sky, it is critical to increase the number of  well-timed MSPs to at least 200 for   detection and full characterization  of the stochastic GW background (spectrum and isotropy).  Gaps in sky coverage yield  
 nonuniform GW sensitivity by a factor of about seven \citep[][]{2018arXiv181211585A}.   This can be rectified by targeting  MSP surveys in particular sky locations.
 Continued surveys  are also critical
for discovering extreme relativistic binary systems. 

{\bf Improved calibration and algorithms.}  TOA precision is limited at the tens of ns level by instrumentation, such as the quality of analog-to-digital converters and post-calibration polarization purity, which  alter measured pulse shapes and the resulting TOAs. 
New instrumentation and polarization calibration methods can provide significant improvement. 
Timing precision will also benefit greatly from 
 improved algorithms for radio frequency interference rejection and advances in the correction of ISM delays.

{\bf Improved solar system ephemerides.}
Timing data are analyzed in the frame of the Solar System's barycenter.  Barycentric referencing  is achieved by   removing  appropriate light-travel times estimated from JPL-provided ephemerides.  Their present accuracy is currently a limiting factor in the detection of the stochastic GW background because systematic effects appear at the orbital frequency of Jupiter when different ephemerides are compared \citep[][]{2018ApJ...859...47A}.  Continued space missions will improve the ephemeris and longer timing data sets will provide independent constraints on barycentric corrections. 

{\bf Better distance measurements.}
Interferometric (VLBI)  distances are necessary for distinguishing period derivatives due to Galactic acceleration from those induced by relativistic effects.
This is critical for tests of GR using MSP binaries. Furthermore, if pulsar distances can be determined to within a GW wavelength (i.e. $\sim$parsecs) the ``pulsar term'' can be included in CW searches, enabling Galactic-scale GW interferometry \cite{2012PhRvL.109h1104M} and increasing sensitivity by roughly a factor of $\sqrt{2}$ and dramatically improving localization capability. Distance estimates to binary MSPs can also be improved by observing the optical emission of their companions with e.g. Gaia \cite{2018arXiv181206262M}.

\section{Prospects for 2020--2030}

New discoveries in the next decade will provide more extreme relativistic binaries and additional, high-quality MSPs for gravitational wave detection.   Large scale surveys for pulsars require thousands of hours on large radio telescopes and deployment of additional machine-learning algorithms.  Sustained timing programs on these objects will yield more stringent tests of theories of gravity and detection and characterization of low-frequency gravitational waves.    These GWs will provide unique information on the mergers of supermassive binary black holes and on cosmic strings and other exotic potential sources of GWs.

Fulfillment of NANOGrav's GW program \citep[][]{2018ApJS..235...37A} will be coordinated through  
the International Pulsar Timing Array (IPTA) consortium  \cite{2013CQGra..30v4010M,2016MNRAS.458.1267V}, which pools data from around the world to provide greater GW sensitivity.  These efforts will allow us to fully exploit the promise of low-frequency GW astronomy and 
multi-messenger astrophysics to probe galaxy formation and evolution through cosmic time. 

\clearpage

\bibliographystyle{unsrt}
\bibliography{ng_overview_papers}

\begin{thebibliography}{10}

\bibitem{2018ApJS..235...37A}
Z.~{Arzoumanian}, A.~{Brazier}, S.~{Burke-Spolaor}, S.~{Chamberlin},
  S.~{Chatterjee}, B.~{Christy}, J.~M. {Cordes}, N.~J. {Cornish},
  F.~{Crawford}, H.~{Thankful Cromartie}, K.~{Crowter}, M.~E. {DeCesar}, P.~B.
  {Demorest}, T.~{Dolch}, J.~A. {Ellis}, R.~D. {Ferdman}, E.~C. {Ferrara},
  E.~{Fonseca}, N.~{Garver-Daniels}, P.~A. {Gentile}, D.~{Halmrast}, E.~A.
  {Huerta}, F.~A. {Jenet}, C.~{Jessup}, G.~{Jones}, M.~L. {Jones}, D.~L.
  {Kaplan}, M.~T. {Lam}, T.~J.~W. {Lazio}, L.~{Levin}, A.~{Lommen}, D.~R.
  {Lorimer}, J.~{Luo}, R.~S. {Lynch}, D.~{Madison}, A.~M. {Matthews}, M.~A.
  {McLaughlin}, S.~T. {McWilliams}, C.~{Mingarelli}, C.~{Ng}, D.~J. {Nice},
  T.~T. {Pennucci}, S.~M. {Ransom}, P.~S. {Ray}, X.~{Siemens}, J.~{Simon},
  R.~{Spiewak}, I.~H. {Stairs}, D.~R. {Stinebring}, K.~{Stovall}, J.~K.
  {Swiggum}, S.~R. {Taylor}, M.~{Vallisneri}, R.~{van Haasteren}, S.~J.
  {Vigeland}, W.~{Zhu}, and {NANOGrav Collaboration}.
\newblock {The NANOGrav 11-year Data Set: High-precision Timing of 45
  Millisecond Pulsars}.
\newblock {\em \apjs}, 235:37, April 2018.

\bibitem{2013CQGra..30v4002C}
J.~M. {Cordes}.
\newblock {Limits to PTA sensitivity: spin stability and arrival time precision
  of millisecond pulsars}.
\newblock {\em Classical and Quantum Gravity}, 30(22):224002, November 2013.

\bibitem{cd85}
J.~M. {Cordes} and G.~S. {Downs}.
\newblock {JPL pulsar timing observations. III - Pulsar rotation fluctuations}.
\newblock {\em \apjs}, 59:343--382, November 1985.

\bibitem{FDJitter}
M.~T. {Lam}, M.~A. {McLaughlin}, Z.~{Arzoumanian}, H.~{Blumer}, P.~R. {Brook},
  H.~T. {Cromartie}, P.~B. {Demorest}, M.~E. {DeCesar}, T.~{Dolch}, J.~A.
  {Ellis}, R.~D. {Ferdman}, E.~C. {Ferrara}, E.~{Fonseca}, N.~{Garver-Daniels},
  P.~A. {Gentile}, M.~L. {Jones}, D.~R. {Lorimer}, R.~S. {Lynch}, C.~{Ng},
  D.~J. {Nice}, T.~T. {Pennucci}, S.~M. {Ransom}, R.~{Spiewak}, I.~H. {Stairs},
  K.~{Stovall}, J.~K. {Swiggum}, S.~J. {Vigeland}, and W.~W. {Zhu}.
\newblock {The NANOGrav 12.5 yr Data Set: The Frequency Dependence of Pulse
  Jitter in Precision Millisecond Pulsars}.
\newblock {\em \apj}, 872:193, February 2019.

\bibitem{sc10}
R.~M. {Shannon} and J.~M. {Cordes}.
\newblock {Assessing the Role of Spin Noise in the Precision Timing of
  Millisecond Pulsars}.
\newblock {\em \apj}, 725:1607--1619, December 2010.

\bibitem{NG9EN}
M.~T. {Lam}, J.~M. {Cordes}, S.~{Chatterjee}, Z.~{Arzoumanian}, K.~{Crowter},
  P.~B. {Demorest}, T.~{Dolch}, J.~A. {Ellis}, R.~D. {Ferdman}, E.~{Fonseca},
  M.~E. {Gonzalez}, G.~{Jones}, M.~L. {Jones}, L.~{Levin}, D.~R. {Madison},
  M.~A. {McLaughlin}, D.~J. {Nice}, T.~T. {Pennucci}, S.~M. {Ransom}, R.~M.
  {Shannon}, X.~{Siemens}, I.~H. {Stairs}, K.~{Stovall}, J.~K. {Swiggum}, and
  W.~W. {Zhu}.
\newblock {The NANOGrav Nine-year Data Set: Excess Noise in Millisecond Pulsar
  Arrival Times}.
\newblock {\em \apj}, 834:35, January 2017.

\bibitem{cs10}
J.~M. {Cordes} and R.~M. {Shannon}.
\newblock {A Measurement Model for Precision Pulsar Timing}.
\newblock {\em arXiv e-prints}, October 2010.

\bibitem{LamDMt}
M.~T. {Lam}, J.~M. {Cordes}, S.~{Chatterjee}, M.~L. {Jones}, M.~A.
  {McLaughlin}, and J.~W. {Armstrong}.
\newblock {Systematic and Stochastic Variations in Pulsar Dispersion Measures}.
\newblock {\em \apj}, 821:66, April 2016.

\bibitem{optimalfreq}
M.~T. {Lam}, M.~A. {McLaughlin}, J.~M. {Cordes}, S.~{Chatterjee}, and T.~J.~W.
  {Lazio}.
\newblock {Optimal Frequency Ranges for Submicrosecond Precision Pulsar
  Timing}.
\newblock {\em \apj}, 861:12, July 2018.

\bibitem{sc17}
R.~M. {Shannon} and J.~M. {Cordes}.
\newblock {Modelling and mitigating refractive propagation effects in precision
  pulsar timing observations}.
\newblock {\em \mnras}, 464:2075--2089, January 2017.

\bibitem{2015ApJ...809...41Z}
W.~W. {Zhu}, I.~H. {Stairs}, P.~B. {Demorest}, D.~J. {Nice}, J.~A. {Ellis},
  S.~M. {Ransom}, Z.~{Arzoumanian}, K.~{Crowter}, T.~{Dolch}, R.~D. {Ferdman},
  E.~{Fonseca}, M.~E. {Gonzalez}, G.~{Jones}, M.~L. {Jones}, M.~T. {Lam},
  L.~{Levin}, M.~A. {McLaughlin}, T.~{Pennucci}, K.~{Stovall}, and
  J.~{Swiggum}.
\newblock {Testing Theories of Gravitation Using 21-Year Timing of Pulsar
  Binary J1713+0747}.
\newblock {\em \apj}, 809:41, August 2015.

\bibitem{2005AJ....129.1993M}
R.~N. {Manchester}, G.~B. {Hobbs}, A.~{Teoh}, and M.~{Hobbs}.
\newblock {The Australia Telescope National Facility Pulsar Catalogue}.
\newblock {\em \aj}, 129:1993--2006, April 2005.

\bibitem{1983ApJ...265L..39H}
R.~W. {Hellings} and G.~S. {Downs}.
\newblock {Upper limits on the isotropic gravitational radiation background
  from pulsar timing analysis}.
\newblock {\em \apjl}, 265:L39--L42, February 1983.

\bibitem{2013PhRvD..88f2005M}
C.~M.~F. {Mingarelli}, T.~{Sidery}, I.~{Mandel}, and A.~{Vecchio}.
\newblock {Characterizing gravitational wave stochastic background anisotropy
  with pulsar timing arrays}.
\newblock {\em \prd}, 88(6):062005, September 2013.

\bibitem{2014PhRvD..90f2011M}
C.~M.~F. {Mingarelli} and T.~{Sidery}.
\newblock {Effect of small interpulsar distances in stochastic gravitational
  wave background searches with pulsar timing arrays}.
\newblock {\em \prd}, 90(6):062011, September 2014.

\bibitem{2018JPhCo...2j5002M}
C.~M.~F. {Mingarelli} and A.~B. {Mingarelli}.
\newblock {Proving the short-wavelength approximation in Pulsar Timing Array
  gravitational-wave background searches}.
\newblock {\em Journal of Physics Communications}, 2(10):105002, October 2018.

\bibitem{2014PhRvD..90h2001G}
Jonathan {Gair}, Joseph~D. {Romano}, Stephen {Taylor}, and Chiara M.~F.
  {Mingarelli}.
\newblock {Mapping gravitational-wave backgrounds using methods from CMB
  analysis: Application to pulsar timing arrays}.
\newblock {\em \prd}, 90:082001, Oct 2014.

\bibitem{2008ApJ...685.1304L}
K.~J. {Lee}, F.~A. {Jenet}, and R.~H. {Price}.
\newblock {Pulsar Timing as a Probe of Non-Einsteinian Polarizations of
  Gravitational Waves}.
\newblock {\em \apj}, 685:1304--1319, October 2008.

\bibitem{2012PhRvD..85h2001C}
S.~J. {Chamberlin} and X.~{Siemens}.
\newblock {Stochastic backgrounds in alternative theories of gravity: Overlap
  reduction functions for pulsar timing arrays}.
\newblock {\em \prd}, 85(8):082001, April 2012.

\bibitem{2015PhRvD..92j2003G}
Jonathan~R. {Gair}, Joseph~D. {Romano}, and Stephen~R. {Taylor}.
\newblock {Mapping gravitational-wave backgrounds of arbitrary polarisation
  using pulsar timing arrays}.
\newblock {\em \prd}, 92:102003, Nov 2015.

\bibitem{2018ApJ...859...47A}
Z.~{Arzoumanian}, P.~T. {Baker}, A.~{Brazier}, S.~{Burke-Spolaor}, S.~J.
  {Chamberlin}, S.~{Chatterjee}, B.~{Christy}, J.~M. {Cordes}, N.~J. {Cornish},
  F.~{Crawford}, H.~{Thankful Cromartie}, K.~{Crowter}, M.~{DeCesar}, P.~B.
  {Demorest}, T.~{Dolch}, J.~A. {Ellis}, R.~D. {Ferdman}, E.~{Ferrara}, W.~M.
  {Folkner}, E.~{Fonseca}, N.~{Garver-Daniels}, P.~A. {Gentile}, R.~{Haas},
  J.~S. {Hazboun}, E.~A. {Huerta}, K.~{Islo}, G.~{Jones}, M.~L. {Jones}, D.~L.
  {Kaplan}, V.~M. {Kaspi}, M.~T. {Lam}, T.~J.~W. {Lazio}, L.~{Levin}, A.~N.
  {Lommen}, D.~R. {Lorimer}, J.~{Luo}, R.~S. {Lynch}, D.~R. {Madison}, M.~A.
  {McLaughlin}, S.~T. {McWilliams}, C.~M.~F. {Mingarelli}, C.~{Ng}, D.~J.
  {Nice}, R.~S. {Park}, T.~T. {Pennucci}, N.~S. {Pol}, S.~M. {Ransom}, P.~S.
  {Ray}, A.~{Rasskazov}, X.~{Siemens}, J.~{Simon}, R.~{Spiewak}, I.~H.
  {Stairs}, D.~R. {Stinebring}, K.~{Stovall}, J.~{Swiggum}, S.~R. {Taylor},
  M.~{Vallisneri}, R.~{van Haasteren}, S.~{Vigeland}, W.~W. {Zhu}, and
  {NANOGrav Collaboration}.
\newblock {The NANOGrav 11 Year Data Set: Pulsar-timing Constraints on the
  Stochastic Gravitational-wave Background}.
\newblock {\em \apj}, 859:47, May 2018.

\bibitem{2014ApJ...789..156M}
S.~T. {McWilliams}, J.~P. {Ostriker}, and F.~{Pretorius}.
\newblock {Gravitational Waves and Stalled Satellites from Massive Galaxy
  Mergers at z <= 1}.
\newblock {\em \apj}, 789:156, July 2014.

\bibitem{2016ApJ...826...11S}
J.~{Simon} and S.~{Burke-Spolaor}.
\newblock {Constraints on Black Hole/Host Galaxy Co-evolution and Binary
  Stalling Using Pulsar Timing Arrays}.
\newblock {\em \apj}, 826:11, July 2016.

\bibitem{2016MNRAS.463L...6S}
A.~{Sesana}, F.~{Shankar}, M.~{Bernardi}, and R.~K. {Sheth}.
\newblock {Selection bias in dynamically measured supermassive black hole
  samples: consequences for pulsar timing arrays}.
\newblock {\em \mnras}, 463:L6--L11, November 2016.

\bibitem{2015ARA&A..53...51S}
R.~S. {Somerville} and R.~{Dav{\'e}}.
\newblock {Physical Models of Galaxy Formation in a Cosmological Framework}.
\newblock {\em \araa}, 53:51--113, August 2015.

\bibitem{2013CQGra..30v4015S}
X.~{Siemens}, J.~{Ellis}, F.~{Jenet}, and J.~D. {Romano}.
\newblock {The stochastic background: scaling laws and time to detection for
  pulsar timing arrays}.
\newblock {\em Classical and Quantum Gravity}, 30(22):224015, November 2013.

\bibitem{2016ApJ...819L...6T}
S.~R. {Taylor}, M.~{Vallisneri}, J.~A. {Ellis}, C.~M.~F. {Mingarelli}, T.~J.~W.
  {Lazio}, and R.~{van Haasteren}.
\newblock {Are We There Yet? Time to Detection of Nanohertz Gravitational Waves
  Based on Pulsar-timing Array Limits}.
\newblock {\em \apjl}, 819:L6, March 2016.

\bibitem{2017NatAs...1..886M}
C.~M.~F. {Mingarelli}, T.~J.~W. {Lazio}, A.~{Sesana}, J.~E. {Greene}, J.~A.
  {Ellis}, C.-P. {Ma}, S.~{Croft}, S.~{Burke-Spolaor}, and S.~R. {Taylor}.
\newblock {The local nanohertz gravitational-wave landscape from supermassive
  black hole binaries}.
\newblock {\em Nature Astronomy}, 1:886--892, November 2017.

\bibitem{2017PhRvL.118r1102T}
S.~R. {Taylor}, J.~{Simon}, and L.~{Sampson}.
\newblock {Constraints on the Dynamical Environments of Supermassive Black-Hole
  Binaries Using Pulsar-Timing Arrays}.
\newblock {\em Physical Review Letters}, 118(18):181102, May 2017.

\bibitem{2018arXiv181108826B}
Sarah {Burke-Spolaor}, Stephen~R. {Taylor}, Maria {Charisi}, Timothy {Dolch},
  Jeffrey~S. {Hazboun}, A.~Miguel {Holgado}, Luke~Zoltan {Kelley}, T.~Joseph~W.
  {Lazio}, Dustin~R. {Madison}, Natasha {McMann}, Chiara M.~F. {Mingarelli},
  Alexander {Rasskazov}, Xavier {Siemens}, Joseph~J. {Simon}, and Tristan~L.
  {Smith}.
\newblock {The Astrophysics of Nanohertz Gravitational Waves}.
\newblock {\em arXiv e-prints}, page arXiv:1811.08826, Nov 2018.

\bibitem{2019NatAs...3....8M}
C.~M.~F. {Mingarelli}.
\newblock {Probing supermassive black hole binaries with pulsar timing}.
\newblock {\em Nature Astronomy}, 3:8--10, January 2019.

\bibitem{2018arXiv181211585A}
K.~{Aggarwal}, Z.~{Arzoumanian}, P.~T. {Baker}, A.~{Brazier}, M.~R. {Brinson},
  P.~R. {Brook}, S.~{Burke-Spolaor}, S.~{Chatterjee}, J.~M. {Cordes}, N.~J.
  {Cornish}, F.~{Crawford}, K.~{Crowter}, T.~{Cromartie}, M.~{DeCesar}, P.~B.
  {Demorest}, T.~{Dolch}, J.~A. {Ellis}, R.~D. {Ferdman}, E.~{Ferrara},
  E.~{Fonseca}, N.~{Garver-Daniels}, P.~{Gentile}, J.~S. {Hazboun}, A.~M.
  {Holgado}, E.~A. {Huerta}, K.~{Islo}, R.~{Jennings}, G.~{Jones}, M.~L.
  {Jones}, A.~R. {Kaiser}, D.~L. {Kaplan}, J.~S. {Key}, M.~T. {Lam}, T.~J.~W.
  {Lazio}, L.~{Levin}, D.~R. {Lorimer}, J.~{Luo}, R.~S. {Lynch}, D.~R.
  {Madison}, M.~A. {McLaughlin}, S.~T. {McWilliams}, C.~M.~F. {Mingarelli},
  C.~{Ng}, D.~J. {Nice}, T.~T. {Pennucci}, N.~S. {Pol}, S.~M. {Ransom}, P.~S.
  {Ray}, X.~{Siemens}, J.~{Simon}, R.~{Spiewak}, I.~H. {Stairs}, D.~R.
  {Stinebring}, K.~{Stovall}, J.~{Swiggum}, S.~R. {Taylor}, J.~E. {Turner},
  M.~{Vallisneri}, R.~{van Haasteren}, S.~J. {Vigeland}, and W.~W. {Zhu}.
\newblock {The NANOGrav 11-Year Data Set: Limits on Gravitational Waves from
  Individual Supermassive Black Hole Binaries}.
\newblock {\em arXiv e-prints}, December 2018.

\bibitem{2007ApJ...659L..33S}
N.~{Seto} and A.~{Cooray}.
\newblock {Searching for Primordial Black Hole Dark Matter with Pulsar Timing
  Arrays}.
\newblock {\em \apjl}, 659:L33--L36, April 2007.

\bibitem{2011PhRvD..84d3511B}
S.~{Baghram}, N.~{Afshordi}, and K.~M. {Zurek}.
\newblock {Prospects for detecting dark matter halo substructure with pulsar
  timing}.
\newblock {\em \prd}, 84(4):043511, August 2011.

\bibitem{2012MNRAS.426.1369K}
K.~{Kashiyama} and N.~{Seto}.
\newblock {Enhanced exploration for primordial black holes using pulsar timing
  arrays}.
\newblock {\em \mnras}, 426:1369--1373, October 2012.

\bibitem{2019arXiv190104490D}
J.~A. {Dror}, H.~{Ramani}, T.~{Trickle}, and K.~M. {Zurek}.
\newblock {Pulsar Timing Probes of Primordial Black Holes and Subhalos}.
\newblock {\em arXiv e-prints}, January 2019.

\bibitem{2005ApJ...627L.125J}
F.~A. {Jenet}, T.~{Creighton}, and A.~{Lommen}.
\newblock {Pulsar Timing and the Detection of Black Hole Binary Systems in
  Globular Clusters}.
\newblock {\em \apjl}, 627:L125--L128, July 2005.

\bibitem{2012ApJ...752...67K}
B.~{Kocsis}, A.~{Ray}, and S.~{Portegies Zwart}.
\newblock {Mapping the Galactic Center with Gravitational Wave Measurements
  Using Pulsar Timing}.
\newblock {\em \apj}, 752:67, June 2012.

\bibitem{2017PhRvD..96l3016M}
D.~R. {Madison}, D.~F. {Chernoff}, and J.~M. {Cordes}.
\newblock {Pulsar timing perturbations from Galactic gravitational wave bursts
  with memory}.
\newblock {\em \prd}, 96(12):123016, December 2017.

\bibitem{2008ARA&A..46..541K}
M.~{Kramer} and I.~H. {Stairs}.
\newblock {The Double Pulsar}.
\newblock {\em \araa}, 46:541--572, September 2008.

\bibitem{2016ApJ...829...55W}
J.~M. {Weisberg} and Y.~{Huang}.
\newblock {Relativistic Measurements from Timing the Binary Pulsar PSR
  B1913+16}.
\newblock {\em \apj}, 829:55, September 2016.

\bibitem{2010Natur.467.1081D}
P.~B. {Demorest}, T.~{Pennucci}, S.~M. {Ransom}, M.~S.~E. {Roberts}, and
  J.~W.~T. {Hessels}.
\newblock {A two-solar-mass neutron star measured using Shapiro delay}.
\newblock {\em \nat}, 467:1081--1083, October 2010.

\bibitem{2013Sci...340..448A}
J.~{Antoniadis}, P.~C.~C. {Freire}, N.~{Wex}, T.~M. {Tauris}, R.~S. {Lynch},
  M.~H. {van Kerkwijk}, M.~{Kramer}, C.~{Bassa}, V.~S. {Dhillon}, T.~{Driebe},
  J.~W.~T. {Hessels}, V.~M. {Kaspi}, V.~I. {Kondratiev}, N.~{Langer}, T.~R.
  {Marsh}, M.~A. {McLaughlin}, T.~T. {Pennucci}, S.~M. {Ransom}, I.~H.
  {Stairs}, J.~{van Leeuwen}, J.~P.~W. {Verbiest}, and D.~G. {Whelan}.
\newblock {A Massive Pulsar in a Compact Relativistic Binary}.
\newblock {\em Science}, 340:448, April 2013.

\bibitem{2018Natur.559...73A}
A.~M. {Archibald}, N.~V. {Gusinskaia}, J.~W.~T. {Hessels}, A.~T. {Deller},
  D.~L. {Kaplan}, D.~R. {Lorimer}, R.~S. {Lynch}, S.~M. {Ransom}, and I.~H.
  {Stairs}.
\newblock {Universality of free fall from the orbital motion of a pulsar in a
  stellar triple system}.
\newblock {\em \nat}, 559:73--76, July 2018.

\bibitem{2014JPhG...41f3101R}
J.~I. {Read}.
\newblock {The local dark matter density}.
\newblock {\em Journal of Physics G Nuclear Physics}, 41(6):063101, June 2014.

\bibitem{2018arXiv180107847K}
K.~{Kashiyama} and M.~{Oguri}.
\newblock {Detectability of Small-Scale Dark Matter Clumps with Pulsar Timing
  Arrays}.
\newblock {\em arXiv e-prints}, January 2018.

\bibitem{2000PhRvL..85.1158H}
W.~{Hu}, R.~{Barkana}, and A.~{Gruzinov}.
\newblock {Fuzzy Cold Dark Matter: The Wave Properties of Ultralight
  Particles}.
\newblock {\em Physical Review Letters}, 85:1158--1161, August 2000.

\bibitem{2017PhRvD..95d3541H}
L.~{Hui}, J.~P. {Ostriker}, S.~{Tremaine}, and E.~{Witten}.
\newblock {Ultralight scalars as cosmological dark matter}.
\newblock {\em \prd}, 95(4):043541, February 2017.

\bibitem{2018PhRvD..98j2002P}
N.~K. {Porayko}, X.~{Zhu}, Y.~{Levin}, L.~{Hui}, G.~{Hobbs}, A.~{Grudskaya},
  K.~{Postnov}, M.~{Bailes}, N.~D.~R. {Bhat}, W.~{Coles}, S.~{Dai},
  J.~{Dempsey}, M.~J. {Keith}, M.~{Kerr}, M.~{Kramer}, P.~D. {Lasky}, R.~N.
  {Manchester}, S.~{Os{\l}owski}, A.~{Parthasarathy}, V.~{Ravi}, D.~J.
  {Reardon}, P.~A. {Rosado}, C.~J. {Russell}, R.~M. {Shannon}, R.~{Spiewak},
  W.~{van Straten}, L.~{Toomey}, J.~{Wang}, L.~{Wen}, X.~{You}, and {PPTA
  Collaboration}.
\newblock {Parkes Pulsar Timing Array constraints on ultralight scalar-field
  dark matter}.
\newblock {\em \prd}, 98(10):102002, November 2018.

\bibitem{2018MNRAS.481.5501C}
R.~N. {Caballero}, Y.~J. {Guo}, K.~J. {Lee}, P.~{Lazarus}, D.~J. {Champion},
  G.~{Desvignes}, M.~{Kramer}, K.~{Plant}, Z.~{Arzoumanian}, M.~{Bailes}, C.~G.
  {Bassa}, N.~D.~R. {Bhat}, A.~{Brazier}, M.~{Burgay}, S.~{Burke-Spolaor},
  S.~J. {Chamberlin}, S.~{Chatterjee}, I.~{Cognard}, J.~M. {Cordes}, S.~{Dai},
  P.~{Demorest}, T.~{Dolch}, R.~D. {Ferdman}, E.~{Fonseca}, J.~R. {Gair},
  N.~{Garver-Daniels}, P.~{Gentile}, M.~E. {Gonzalez}, E.~{Graikou},
  L.~{Guillemot}, G.~{Hobbs}, G.~H. {Janssen}, R.~{Karuppusamy}, M.~J. {Keith},
  M.~{Kerr}, M.~T. {Lam}, P.~D. {Lasky}, T.~J.~W. {Lazio}, L.~{Levin},
  K.~{Liu}, A.~N. {Lommen}, D.~R. {Lorimer}, R.~S. {Lynch}, D.~R. {Madison},
  R.~N. {Manchester}, J.~W. {McKee}, M.~A. {McLaughlin}, S.~T. {McWilliams},
  C.~M.~F. {Mingarelli}, D.~J. {Nice}, S.~{Os{\l}owski}, N.~T. {Palliyaguru},
  T.~T. {Pennucci}, B.~B.~P. {Perera}, D.~{Perrodin}, A.~{Possenti}, S.~M.
  {Ransom}, D.~J. {Reardon}, S.~A. {Sanidas}, A.~{Sesana}, G.~{Shaifullah},
  R.~M. {Shannon}, X.~{Siemens}, J.~{Simon}, R.~{Spiewak}, I.~{Stairs},
  B.~{Stappers}, D.~R. {Stinebring}, K.~{Stovall}, J.~K. {Swiggum}, S.~R.
  {Taylor}, G.~{Theureau}, C.~{Tiburzi}, L.~{Toomey}, R.~{van Haasteren},
  W.~{van Straten}, J.~P.~W. {Verbiest}, J.~B. {Wang}, X.~J. {Zhu}, and W.~W.
  {Zhu}.
\newblock {Studying the Solar system with the International Pulsar Timing
  Array}.
\newblock {\em \mnras}, 481:5501--5516, December 2018.

\bibitem{2007MNRAS.382..879S}
E.~R. {Siegel}, M.~P. {Hertzberg}, and J.~N. {Fry}.
\newblock {Probing dark matter substructure with pulsar timing}.
\newblock {\em \mnras}, 382:879--885, December 2007.

\bibitem{2012PhRvL.109h1104M}
C.~M.~F. {Mingarelli}, K.~{Grover}, T.~{Sidery}, R.~J.~E. {Smith}, and
  A.~{Vecchio}.
\newblock {Observing the Dynamics of Supermassive Black Hole Binaries with
  Pulsar Timing Arrays}.
\newblock {\em Physical Review Letters}, 109(8):081104, August 2012.

\bibitem{2018arXiv181206262M}
C.~M.~F. {Mingarelli}, L.~{Anderson}, M.~{Bedell}, and D.~N. {Spergel}.
\newblock {Improving Binary Millisecond Pulsar Distances with Gaia}.
\newblock {\em arXiv e-prints}, December 2018.

\bibitem{2013CQGra..30v4010M}
R.~N. {Manchester} and {IPTA}.
\newblock {The International Pulsar Timing Array}.
\newblock {\em Classical and Quantum Gravity}, 30(22):224010, November 2013.

\bibitem{2016MNRAS.458.1267V}
J.~P.~W. {Verbiest}, L.~{Lentati}, G.~{Hobbs}, R.~{van Haasteren}, P.~B.
  {Demorest}, G.~H. {Janssen}, J.-B. {Wang}, G.~{Desvignes}, R.~N. {Caballero},
  M.~J. {Keith}, D.~J. {Champion}, Z.~{Arzoumanian}, S.~{Babak}, C.~G. {Bassa},
  N.~D.~R. {Bhat}, A.~{Brazier}, P.~{Brem}, M.~{Burgay}, S.~{Burke-Spolaor},
  S.~J. {Chamberlin}, S.~{Chatterjee}, B.~{Christy}, I.~{Cognard}, J.~M.
  {Cordes}, S.~{Dai}, T.~{Dolch}, J.~A. {Ellis}, R.~D. {Ferdman}, E.~{Fonseca},
  J.~R. {Gair}, N.~E. {Garver-Daniels}, P.~{Gentile}, M.~E. {Gonzalez},
  E.~{Graikou}, L.~{Guillemot}, J.~W.~T. {Hessels}, G.~{Jones},
  R.~{Karuppusamy}, M.~{Kerr}, M.~{Kramer}, M.~T. {Lam}, P.~D. {Lasky},
  A.~{Lassus}, P.~{Lazarus}, T.~J.~W. {Lazio}, K.~J. {Lee}, L.~{Levin},
  K.~{Liu}, R.~S. {Lynch}, A.~G. {Lyne}, J.~{Mckee}, M.~A. {McLaughlin}, S.~T.
  {McWilliams}, D.~R. {Madison}, R.~N. {Manchester}, C.~M.~F. {Mingarelli},
  D.~J. {Nice}, S.~{Os{\l}owski}, N.~T. {Palliyaguru}, T.~T. {Pennucci},
  B.~B.~P. {Perera}, D.~{Perrodin}, A.~{Possenti}, A.~{Petiteau}, S.~M.
  {Ransom}, D.~{Reardon}, P.~A. {Rosado}, S.~A. {Sanidas}, A.~{Sesana},
  G.~{Shaifullah}, R.~M. {Shannon}, X.~{Siemens}, J.~{Simon}, R.~{Smits},
  R.~{Spiewak}, I.~H. {Stairs}, B.~W. {Stappers}, D.~R. {Stinebring},
  K.~{Stovall}, J.~K. {Swiggum}, S.~R. {Taylor}, G.~{Theureau}, C.~{Tiburzi},
  L.~{Toomey}, M.~{Vallisneri}, W.~{van Straten}, A.~{Vecchio}, Y.~{Wang},
  L.~{Wen}, X.~P. {You}, W.~W. {Zhu}, and X.-J. {Zhu}.
\newblock {The International Pulsar Timing Array: First data release}.
\newblock {\em \mnras}, 458:1267--1288, May 2016.

\end{thebibliography}

\end{document}